*Article*

# A FPGA-Based Broadband EIT System for Complex Bioimpedance Measurements—Design and Performance Estimation

Roman Kusche [1,2,*], Ankit Malhotra [1,3], Martin Ryschka [1,*], Gunther Ardelt [4], Paula Klimach [1,2] and Steffen Kaufmann [1,3]

[1] Laboratory of Medical Electronics (LME), Lübeck University of Applied Sciences, 23562 Lübeck, Germany; E-Mails: malhotra@imt.uni-luebeck.de (A.M.); paula.klimach@fh-luebeck.de (P.K.); kaufmann@imt.uni-luebeck.de (S.K.)

[2] Graduate School for Computing in Medicine and Life Sciences, University of Lübeck, 23562 Lübeck, Germany

[3] Institute of Medical Engineering, University of Lübeck, 23562 Lübeck, Germany

[4] Center of Excellence CoSA (Communications—Systems—Applications), Lübeck University of Applied Sciences, 23562 Lübeck, Germany; E-Mail: gunther.ardelt@fh-luebeck.de

**\*** Authors to whom correspondence should be addressed; E-Mails: roman.kusche@fh-luebeck.de (R.K.); martin.ryschka@fh-luebeck.de (M.R.); Tel.: +49-451-300-5400 (R.K.); +49-451-300-5021 (M.R.).

Academic Editor: Ignacio Bravo-Muñoz

*Received: 5 May 2015 / Accepted: 22 July 2015 / Published: 29 July 2015*

**Abstract:** Electrical impedance tomography (EIT) is an imaging method that is able to estimate the electrical conductivity distribution of living tissue. This work presents a field programmable gate array (FPGA)-based multi-frequency EIT system for complex, time-resolved bioimpedance measurements. The system has the capability to work with measurement setups with up to 16 current electrodes and 16 voltage electrodes. The excitation current has a range of about 10 µA to 5 mA, whereas the sinusoidal signal used for excitation can have a frequency of up to 500 kHz. Additionally, the usage of a chirp or rectangular signal excitation is possible. Furthermore, the described system has a sample rate of up to 3480 impedance spectra per second (ISPS). The performance of the EIT system is demonstrated with a resistor-based phantom and tank phantoms. Additionally, first measurements taken from the human thorax during a breathing cycle are presented.

## 1. Introduction

In the past bioimpedance measurements became a popular method to determine the characteristics of tissue [1], *i.e.*, a known application is the determination of body composition [2]. By performing time-resolved measurements, it is further possible to measure the impedance changes that happen, for example, during the arrival of a pulse wave [3]. With knowledge of the bioimpedance of various points around an object under testing, the internal conductivity distribution can be estimated. The method is known as electrical impedance tomography (EIT) [4]. First clinical applications of EIT are respiration monitoring [5] and breast cancer detection [6,7].

To reconstruct an image of the conductivity distributions, different transfer-impedance measurements are necessary. Since the impedances change over time, it is fundamental to measure all needed transfer-impedances for a single image in a short fraction of time. Instead of implementing identical measurement circuits for each of the transfer-impedances, just one circuit is realized. By using multiplexers, all transfer-impedances of interest can be measured successively. Since the measurement of a complete single image is much faster than the expected impedance changes, the influence of the resulting time lags can be neglected. The current research represents this kind of a serial EIT-system.

Common multi-frequency EIT systems like KHU Mark2 [8], KHU Mark2.5 [9], UCLH Mk 2.5 EIT [10], *ACT 4* [11], or the system of the Dartmouth College research group [12] are able to measure the impedances' magnitudes and phases over time using usually one chosen excitation frequency or the addition of a few sinusoidal waves with different frequencies. However, the introduced EIT system measures the complete spectrum between about 10 kHz and 380 kHz for each transfer impedance over time, achieving an additional dimension of information for the subsequent image reconstruction.

This is an enhancement of a previously published FPGA based single-channel Bioimpedance Measurement System (BMS), developed by the authors' group [3]. By using a simple multiplexing algorithm it is possible to realize up to 16 current and 16 voltage channels with four 8-to-1 multiplexers. In addition to the single frequency measurement mode, the system is able to use chirp signal excitation [13] and to determine the magnitude as well as the phase of the impedances. Depending on the multiplexing algorithm, the system allows measurements of up to 3480 impedance spectra per second (ISPS).

Using all current and voltage channels, an image refresh rate of about 4 images per second (IPS) is achievable, which is fast enough, for example, to record respiration-caused impedance changes. Without executing reciprocal measurements, the image refresh rate can be increased to up to 8 IPS.

In most measurement ranges, the overall uncertainties were measured to be below 1% for the impedance magnitude. This work describes the developed EIT system as well as its verification. Moreover, some exemplary measurements with phantoms and over the thorax of a healthy subject are presented.

## 2. Materials and Methods

*2.1. System Architecture*

The block diagram in Figure 1 represents the measurement system, which works like an interface between the object under testing and the host PC. Its tasks are excitation signal generation, multiplexing, data acquisition, and signal pre-processing, with the principle of impedance measurement is based on [3]. The host PC is used for further signal processing and display of the measurement results.

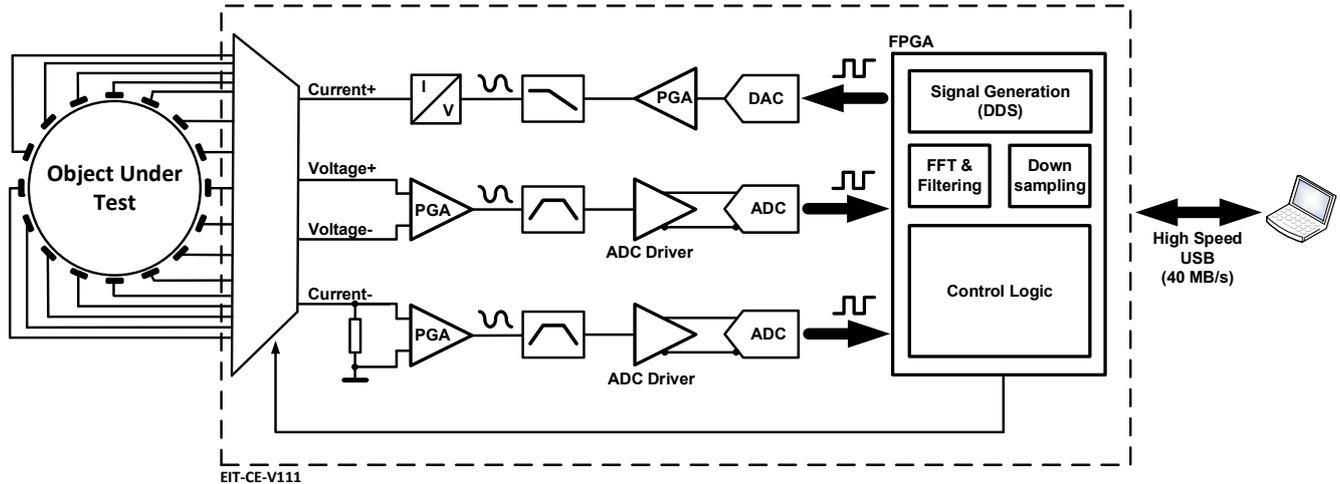

**Figure 1.** Principle block diagram of the developed FPGA-based broadband electrical impedance tomography (EIT) system. The system generates the excitation current, measures the voltage drops across the transfer impedances, and is able to change the electrode configurations via multiplexers. The measurement data is sent via a high speed USB link to a host PC.

The object under testing is connected to the EIT system's multiplexers via electrodes and measurement cables with driven shields. The excitation signal is generated by the FPGA's internal direct digital synthesis (DDS), which works with the same 50 MHz clock as the following 16-bit digital-to-analogue converter (DAC, LTC1668 from Linear Technology) to minimize clock jitter-induced measurement errors. The generated excitation signal can be amplified with programmable gain amplifiers (PGA, AD8251 from Analog Devices) in four steps (G = {1; 2; 4; 8}). Afterwards, the amplified signal passes a passive first order reconstruction filter with a cut-off frequency of $f_c$ = 1.7 MHz for reducing out-of-band noise. The excitation current is subsequently generated via a voltage-controlled current source (VCCS). This current source is based on an AD8130 differential amplifier (Analog Devices) [14]. By changing both the PGA and the DDS configuration, up to 160 different current amplitudes in a range of about 10 μA to 5 mA can be adjusted. The available frequency range of the excitation signal is between approximately 10 kHz and 500 kHz. In addition to sinusoidal excitations, the DDS can be used to generate rectangular signals and chirp signals. In the current setup, the chirp signal has a frequency range of $f_{start}$ = 12 kHz to $f_{stop}$ = 378.625 kHz, with an excitation period of $T_{exec}$ = 40.96 μs. The excitation current, which flows through the unknown impedance via the current electrodes (selected via the multiplexer) is measured via a shunt resistor.

The voltage drop across the shunt resistor, as well as the voltage drop across the unknown impedance of interest, are measured in the same way. Both voltages are amplified by PGAs, to adapt the amplitudes to the ideal input range of the following stages. Each signal is filtered by two active filters. The first filter is a second order high pass with a cut-off frequency of $f_{c,HP}$ = 1 kHz, which is implemented in Sallen–Key topology. The second filter is a low pass of third order with a cut-off frequency of $f_{c,LP}$ = 2 MHz, implemented in multiple feedback topology. These topologies and cut-off frequencies are chosen to minimize distortions in the signal frequency range of about 10 kHz to 500 kHz. After filtering, the signals are digitized by a dual-channel 14 bit analogue-to-digital converter (ADC, LTC2296 from Linear Technology) with a sample rate of 25 mega samples per second (MSPS). The clock of the ADC is coherent with the DDS and the DAC clock.

In the next processing step, the FPGA decimates the digitized signals. Before the voltage and current data of the impedance measurement is fed into two 1024-point fast Fourier transformations (FFT) for calculating both complex spectra, an optional averaging can be performed. The current version of the EIT system measures up to 2 × 3480 complex spectra per second. The resulting data are transmitted via a Universal Serial Bus (USB) interface chip (FT2232HL from Future Technology Devices International) to the host computer for further signal processing. Using MathWorks MATLAB, the impedance spectra are calculated by complex divisions of the received voltage and current spectra. Additionally, it is possible to change the measurement parameters such as excitation waveform and PGA settings with the host PC to provide further flexibility. The reconstruction is performed with the open source MATLAB framework EIDORS (http://www.eidors.org) [15].

To minimize the influences of systematic errors and tolerances of the electrical components, an optional calibration of the system can be done. The resulting data can be involved in the calculation of the impedances.

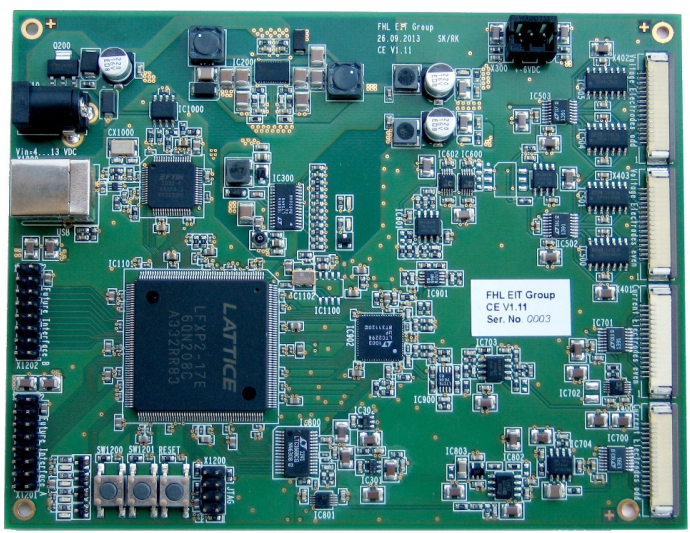

**Figure 2.** Populated printed circuit board (PCB) of the described electrical impedance tomography (EIT) system. The PCB has a size of approximately 142 mm × 110 mm and contains more than 400 components. In addition to the parts described, it has two more interfaces for future use purposes like synchronous ECG measurements.

The system is powered by a 5 $V_{DC}$ medical power supply (SINPRO model no MPU31-102). The system generates ±6 $V_{DC}$ internally for the analogue components, as well as 3.3 $V_{DC}$ and 1.2 $V_{DC}$ for the digital parts through a DC-DC Converter (based on a LT3640 from Linear Technology). Furthermore, for electrical safety considerations it is recommend to insulate the USB via a fiber-optic USB hub. Figure 2 shows an image of the realized printed circuit board (PCB) without housing and cables. Further information about the bioimpedance measurement principle is published in [3].

*2.2. Current and Voltage Channel Multiplexing*

To keep the multiplexing efficient, a simple architecture is used that separates current and voltage electrodes from each other. With this kind of architecture, only the combination of even numbered electrodes together with odd numbered electrodes is realized (see Figure 3). Due to this simplification, the number of required multiplexers is reduced to four 8-to-1 multiplexers for 16 current and 16 voltage channels.

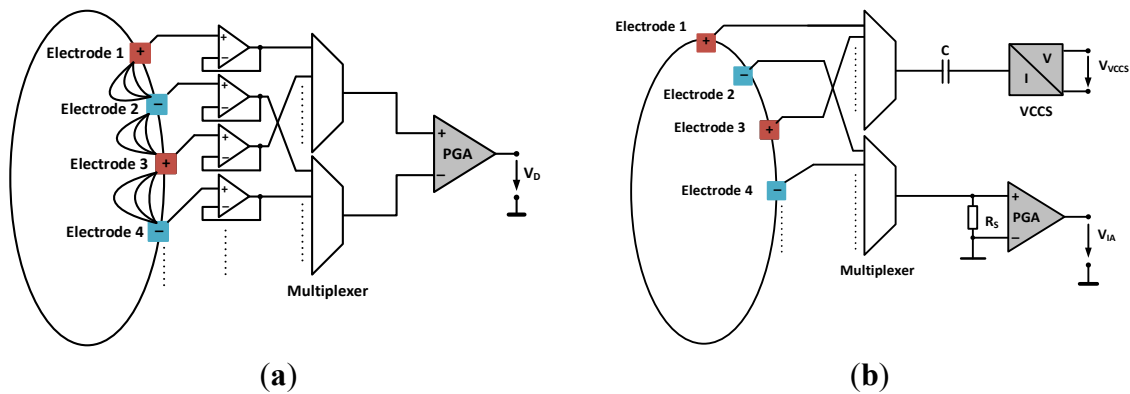

(**a**) (**b**)

**Figure 3.** Multiplexing principle for voltage measurements (**a**) and current injection (**b**).

To avoid the influence of the multiplexers' input capacities, the voltage channels are decoupled by buffer amplifiers (OPA4134 from Texas Instruments). The DC voltage offsets at the electrodes, caused by the electrode–skin interfaces, are not reduced by the operation amplifiers since they are realized as buffer amplifiers. Assuming that the occurring DC offsets at all electrodes are approximately in the same range, the major error is eliminated by the subtraction of the signals in the PGA. Caused by the high common mode rejection ratio (CMRR ≈ 80 dB at f =50 kHz, G = 4) of the chosen PGA, the influence of the common mode can be neglected.

During each switching process the measurement is interrupted until the filters and multiplexers are in a steady-state again. Afterwards, the current and voltage spectra are measured and transmitted to the host PC. A frame consists of 16 current settings and 13 voltage measurements for each current setting. The system is able to measure all of these resulting 208 channels in 250 ms. Thus the resulting frame rate is 4 IPS. By relinquishing reciprocal measurements the frame rate can be increased up to 8 IPS.

*2.3. Software Architecture of the EIT System*

The developed software architecture is divided into different parts: the firmware present on the EIT system hardware as well as interface software and a framework that is based on MathWorks MATLAB running on the host PC. The signal transmission is bidirectional. Firstly, for initialization, the configuration data from the user interface is transferred to the embedded system. Secondly, the acquired measurement data is transmitted from the EIT system to the user interface. Figure 4 illustrates the developed software architecture and the information flow between the measurement device and the user interface.

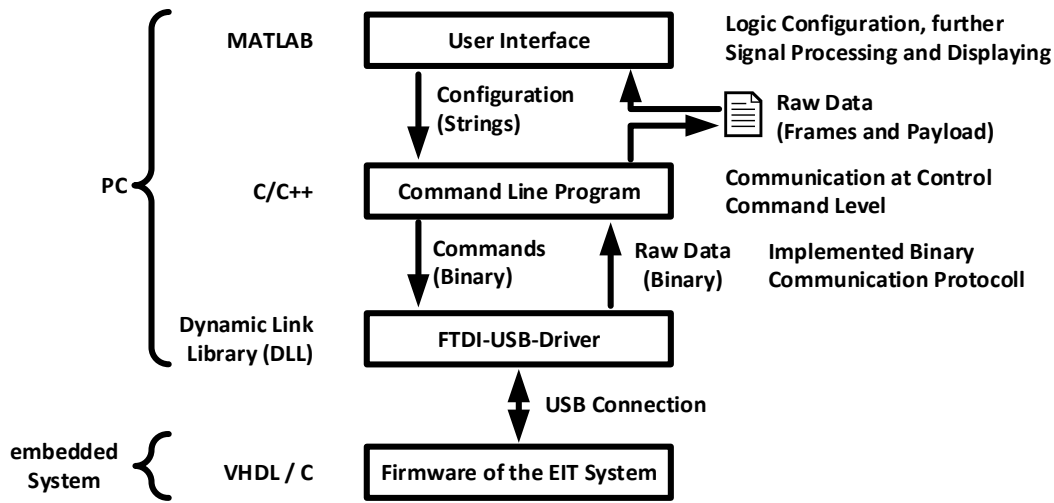

**Figure 4.** Software architecture and information flow of the EIT system.

The implemented user interface provides settings for configuration and signal processing and the displays of the measurement results. Additionally it provides payload scaling and various error detections and it converts the logical commands into strings for the command line program. To keep the data exchange between both operating system processes (MATLAB and the command lines program) simple, a file for reading and writing raw data is used. The user interface reads the raw data from this file, parses it, and extracts the payload afterwards. In order to keep the configuration interface between the user interface and the command line program simple, the commands are capsulated before transmitting. The command line program is used to control the USB interface. It integrates the Dynamic-Link-Library (DLL), which is provided by the manufacturer of the USB-Interface chip.

The firmware of the EIT system consists of a Very High Speed Integrated Circuit Hardware Description Language (VHDL) part and C program code for an implemented 8-bit softcore microcontroller (Mico8 from Lattice Semiconductor) present inside the FPGA.

*2.4. System Performance*

For methodical verification, various calibration resistors $R_{Cal}$ are connected via an interface board to the current and voltage channels using the four-wire technique. Afterwards, impedance measurements with all channels are executed to acquire the current and voltage spectra using sinusoidal and chirp excitation.

The component values used are checked by a Wayne Kerr Precision Component Analyzer 6425 at a frequency of 1 kHz in high level accuracy mode with an uncertainty of less than 0.1%.

The electrical contact between the interface board and calibration resistor boards (Figure 5b) is realized with gold plated contacts at which the contact impedances $Z_C$ can be neglected due to the four-wire technique. Figure 5 shows the principle circuit and an image of the interface board with and without a calibration resistor board.

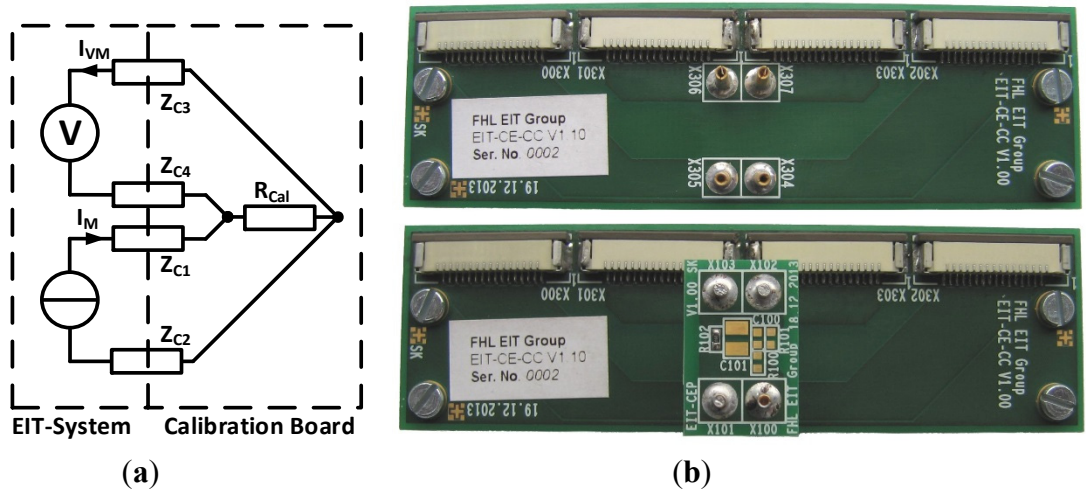

**Figure 5.** Principle circuit of measurement channels (**a**) and interface board (**b**) with and without calibration resistor board.

To determine the system performance, the signal to noise and distortion ratio (SINAD), the effective number of bits (ENOB), the spurious free dynamic range (SFDR), and the total harmonic distortion+noise (THD+N) are calculated from the measurement results. The measurements were performed with known resistors and excitation signals. Figure 6 shows the resulting current and voltage spectra with the highest and lowest magnitude of all channels, using sinusoidal and chirp excitations on a 46.5 Ω resistor. The results are averages of 100 measurements of each 208 channels.

The measurements are done with the interface board (see Figure 5) and with an excitation current of 5 mA. Table 1 shows the resulting SINAD, ENOB, THD+N, and SFDR values of the averaged voltage and current spectra in a frequency range of 48.8 kHz to 391 kHz. The worst ENOB values (14.0 bit for the voltage spectrum, 14.1 bit for the current spectrum) for the chosen measurement range lead to an estimated relative resolution of 86 ppm, corresponding to an absolute resolution of 4.5 mΩ in the case of sinusoidal excitation signals. Chirp excitation yields a relative resolution, depending on the particular spectral energy, of approx. 280 ppm to 847 ppm, corresponding to an absolute resolution of 13.7 mΩ to 41.5 mΩ over the specified frequency range.

(a) Voltage spectrum with sinusoidal excitation of f ≈ 48.8 kHz. The achieved SINAD is ≈89 dBFS, the ENOB is ≈14.4 bit, the SFDR is ≈67 dB and the THD+N is ≈56 dB.
(b) Current spectrum with sinusoidal excitation of f ≈ 48.8 kHz. The achieved SINAD is ≈89 dBFS, ENOB is ≈14.5 bit, the SFDR is ≈69 dB and the THD+N is ≈59 dB.
(c) Voltage spectrum with chirp excitation with a period of T = 40.96 μs.
(d) Current spectrum with chirp excitation with a period of T = 40.96 μs.

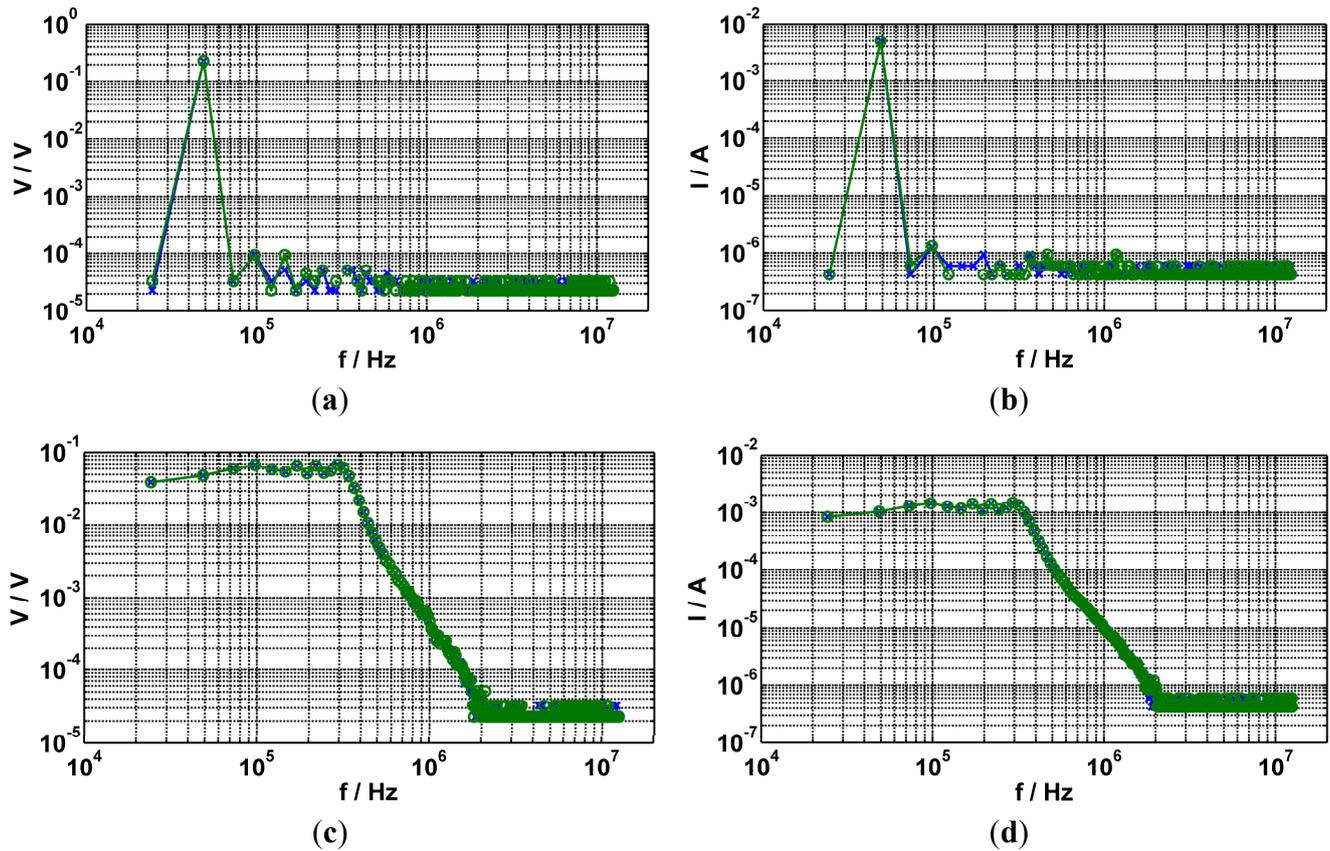

**Figure 6.** SINAD measurements with a 46.5-Ω resistor using the optimal measurement range at $f_s$ = 25 MHz. The spectra with the highest and the lowest magnitude of 20,800 single measurements (100 per channel) are shown. The employed excitation current is 5 mA.

**Table 1.** SINAD, ENOB, SFDR, THD+N for the voltage and current spectra at different frequencies based on 20,800 measurements taken on a 46.5-Ω resistor with a sinusoidal excitation current of 5 mA. The values are calculated using the spectra with the highest and the lowest magnitude.

| f | 48.8 kHz | 97.7 kHz | 195 kHz | 391 kHz |
|---|---|---|---|---|
| $SINAD_V/dB_{FS}$ | 89.0/88.6 | 87.9/88.0 | 87.5/88.1 | 87.4/86.1 |
| $SINAD_I/dB_{FS}$ | 89.5/88.7 | 87.9/87.7 | 87.9/87.6 | 86.6/86.8 |
| $ENOB_V$/bit | 14.5/14.4 | 14.3/14.3 | 14.2/14.3 | 14.2/14.0 |
| $ENOB_I$/bit | 14.6/14.4 | 14.3/14.3 | 14.3/14.3 | 14.2/14.1 |
| $SFDR_V$/dB | 69.0/67.1 | 64.4/64.4 | 57.6/56.5 | 49.9/49.6 |
| $SFDR_I$/dB | 68.5/69.0 | 64.9/63.3 | 59.1/59.7 | 54.2/54.6 |
| $THD_V+N_V$/dB | 57.4/56.6 | 57.2/57.8 | 54.2/53.6 | 48.5/48.0 |
| $THD_I+N_I$/dB | 58.9/59.0 | 59.7/59.5 | 58.1/58.1 | 53.5/53.8 |

*2.5. Channel-Dependent Deviations*

Each channel-dependent deviation of the EIT system is determined by carrying out four-wire measurements with different resistors and comparing the results of all the channels with each other. This measurement is again executed with the interface board shown in Figure 5. The results prove that channel calibrations are not necessary because the statistical errors are in the same range as the systematic differences between the channels. Figure 7 shows the result of an exemplary measurement of a 46.57-Ω resistor. All 208 possible channels are measured 100 times with a 48.8 kHz sinusoidal excitation current of 5 mA. Each of these 20,800 impedance measurements is in the range of 46.265 Ω ± 0.2‰. Due to these very small channel-dependent deviations, there is no need to calibrate all channels separately. Therefore, a calibration can be executed with just one channel.

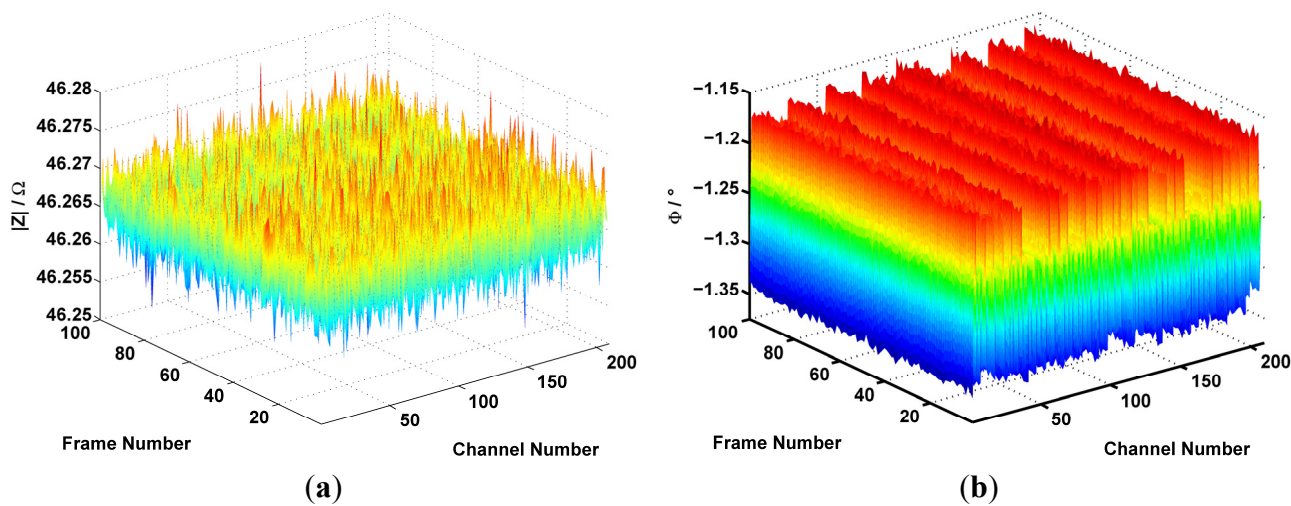

(**a**)    (**b**)

**Figure 7.** Measurement of the channel-dependent magnitude (**a**) and phase (**b**) deviations with a 46.57-Ω resistor using a sinusoidal 48.8 kHz excitation current of 5 mA. One hundred frames from each of the 208 channels are acquired.

*2.6. Verification with an R-R||C Phantom*

After demonstrating that channel-based calibration of the EIT system is not necessary, the accuracy of the system is verified with a complex impedance phantom. For this, the transfer impedances of the phantom are measured and afterwards a comparison with the simulation results is achieved through LTspiceIV from Linear Technology (http://www.linear.com/designtools/software/#LTspice).

For the verification of magnitude and phase, a phantom as shown in Figure 8 with component values of $R_S$ = 19.9 Ω, $R_P$ = 19.86 Ω, and $C_P$ = 99.7 pF is used (Measured with a Wayne Kerr Precision Component Analyzer 6425 at 1 kHz in high level accuracy mode with an uncertainty of less than 0.1%). These values are chosen to provoke typical voltage ranges of a thorax impedance measurement as well as significant phase shifts of about 20° [16]. The impedances are measured with adjacent electrodes and an excitation current of 5 mA in a frequency range of up to 391 kHz.

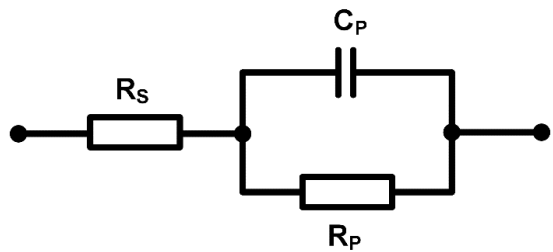

**Figure 8.** R-R||C phantom to verify the EIT system.

In Figure 9a the results of all 208 transfer impedance measurements together with the theoretical values for magnitude and phase are given. Figure 9b shows the calculated deviations of the measured magnitudes and phases. The magnitude has a relative deviation of less than 0.75% over the whole frequency range; the absolute phase deviation increases significantly with the frequency. Since the phase shifts are constant and typical for every channel, there is no influence on image reconstruction because of measuring just phase changes depending on time. Additionally, the system can be calibrated.

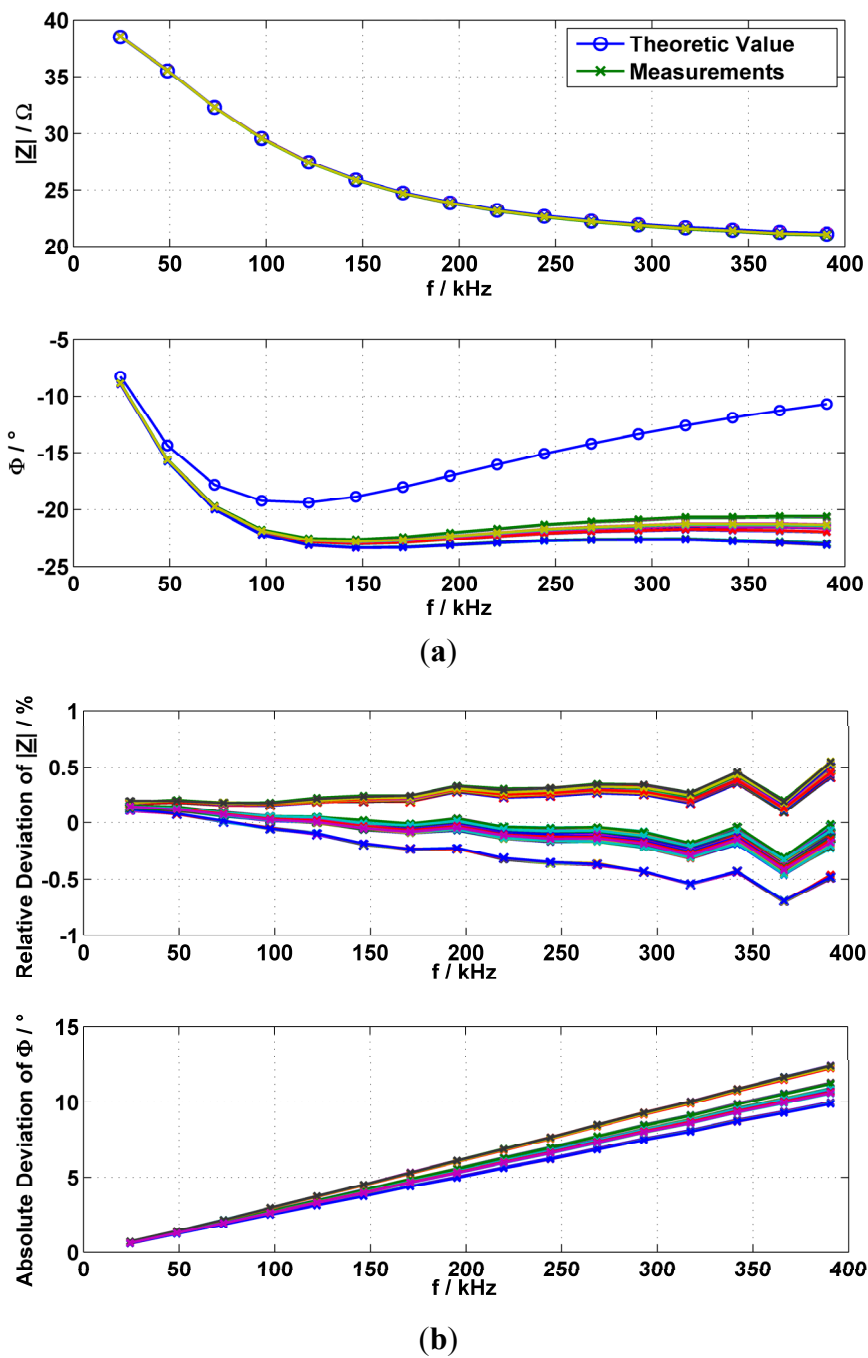

**Figure 9.** Measured impedances divided into magnitude and phases in comparison with the theoretic values. (**a**) Measured magnitude and phase compared to the theoretic values of the R-R||C phantom; (**b**) relative and absolute deviations of measured compared to theoretic values of the R-R||C phantom.

## 3. Results

*3.1. Micro-Tank Measurements*

For demonstration purposes, the first measurements with the EIT system are acquired with a micro-tank phantom based on [17,18]. The major advantage is that the phantom is very compact, but still represents the approximate behavior of larger phantoms. The basic idea is to utilize the milled golden vertical interconnect accesses (VIA) of a PCB as micro electrodes. Figure 10a shows the top view of the principal setup of the micro tank and an image of an electrode PCB with 16 golden micro electrodes (Figure 10b). The realization of the electrodes has a crucial influence on the electrode's contact, hence the manufacturing tolerances have to be very low.

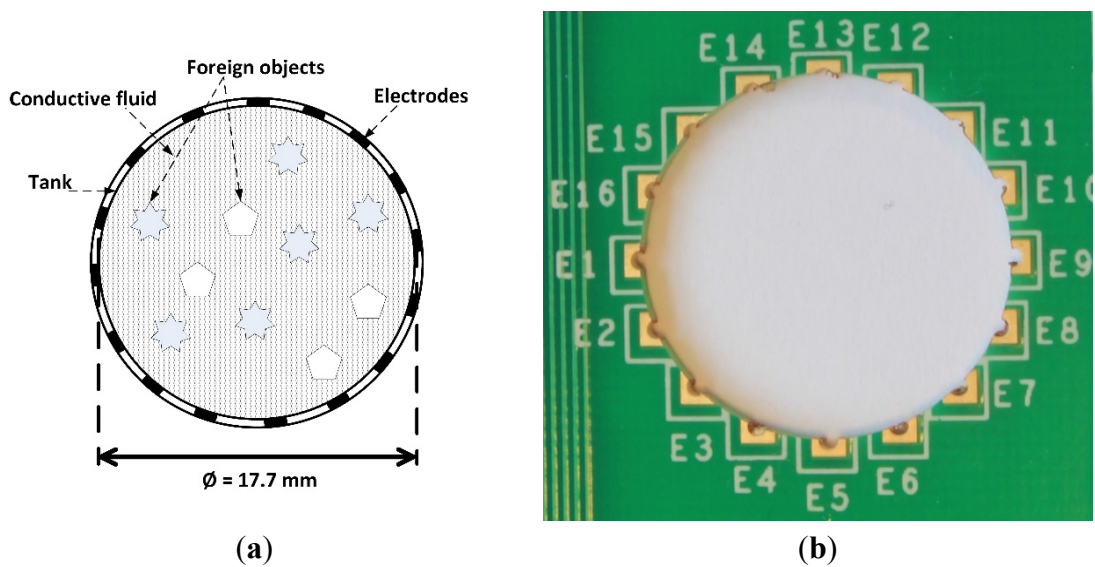

(**a**) (**b**)

**Figure 10.** Principle drawing of the micro-tank phantom used (**a**) and image of an electrode PCB (**b**).

Figure 11 shows the first reconstruction results achieved with the developed EIT system. For the measurement, the micro-tank is filled with a homogeneous agarose gel (relative weights: 0.6% agarose, 0.1% sodium chloride, and 99.3% distilled water), which has a specific resistance of about 4.55 Ωm. Afterwards a solution of sodium chloride (0.9% sodium chloride, 99.1% distilled water) is injected into the agarose gel to produce a small salt water ball. The agarose gel grants a good positioning inside the micro-tank phantom. By acquiring the transfer impedances before and after the injection, the relative conductivity changes were calculated with EIDORS.

The results demonstrate the essential functionality of the EIT system. Due to the high contact impedances of the electrodes, which are caused by their small geometry, the influence of manufacturing tolerances, and the limited repeatability of the measurement scenarios, a larger tank phantom is preferable and is used afterwards.

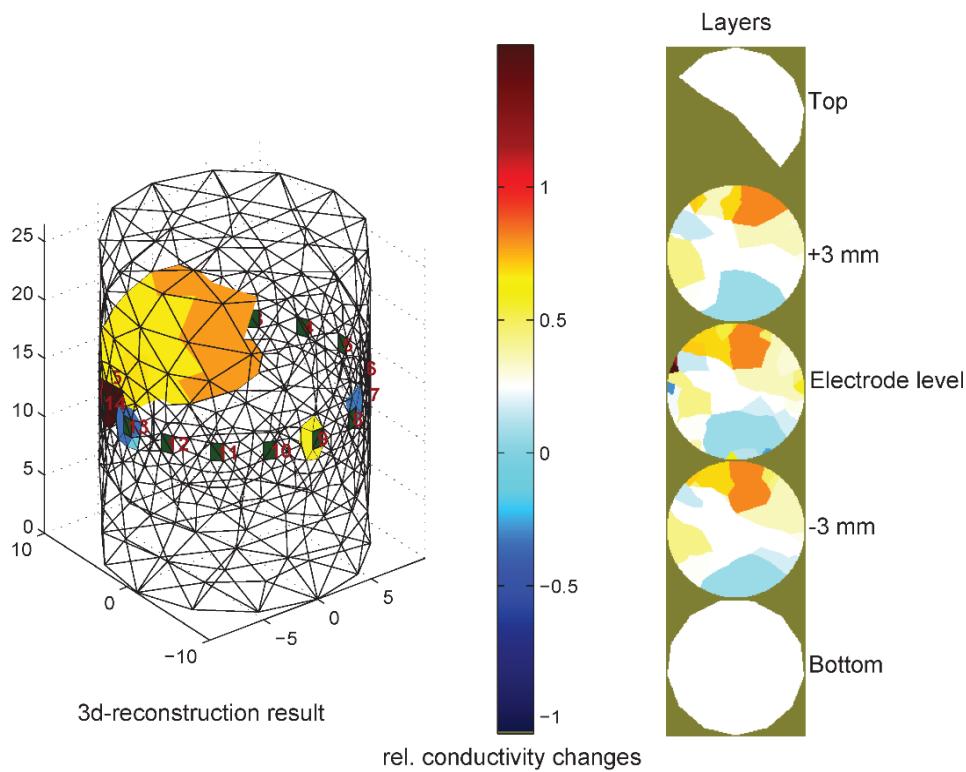

**Figure 11.** EIDORS reconstruction result of the conductivity changes inside the built micro-tank.

*3.2. Tank Measurements*

Because of the described electrode problems of the micro-tank, a larger tank filled with a solution of sodium chloride is used for measurements. This tank phantom consists of 4 mm thick Plexiglas. It has an inner diameter of 242 mm and a height of 300 mm.

There are three rings of electrodes in the phantom with distances of 65 mm between them, consisting of 16 compound electrodes [19] made of stainless steel (X2CrNiMo17-12-2, V4A) with a thickness of 0.5 mm. The inner electrodes have a diameter of 10 mm and the outer electrodes have an inner diameter of 20 mm and an outer diameter of 40 mm. For this particular research, the outer electrode rings are not connected to the EIT system. The connection of the measurement cables at the electrodes is realized with stainless steel screws (M3), which are screwed through the Plexiglas and sealed by silicone. The positioning of the measurement objects in the tank is accomplished with a phantom holder. Figure 12 shows the tank in combination with the phantom holder inclusive of two phantoms. Both the holder and the phantoms were designed by SolidWorks (from Dassault Systèmes). This construction allows very accurate positioning of the phantoms.

The vertical adjustment is realized by a threaded bar (M4) made of stainless steel. All other components of the holder as well as the phantoms consist of Acrylnitril-Butadien-Styrol (ABS) plastic material and were printed with a 3D printer (MakerBot Replicator 2X).

For the measurements, the tank was filled with 11.7 L of tap water mixed with 20 g of sodium chloride to decrease the resistance to approximately 2.7 Ωm. The EIT system and the electrodes are connected via coaxial cables and an interface PCB, which provides, in addition to plug connectors, driven shields (the principle circuit diagram can be found in [3]) for the measurement cables. The measurement cables have a length of 1.5 m and are made of RG174 coaxial cables. On the interface board the cables are connected via BNC-connectors and with the electrodes via alligator clips. The power supply of the interface board is provided by the EIT system. Figure 13 shows the complete measurement setup.

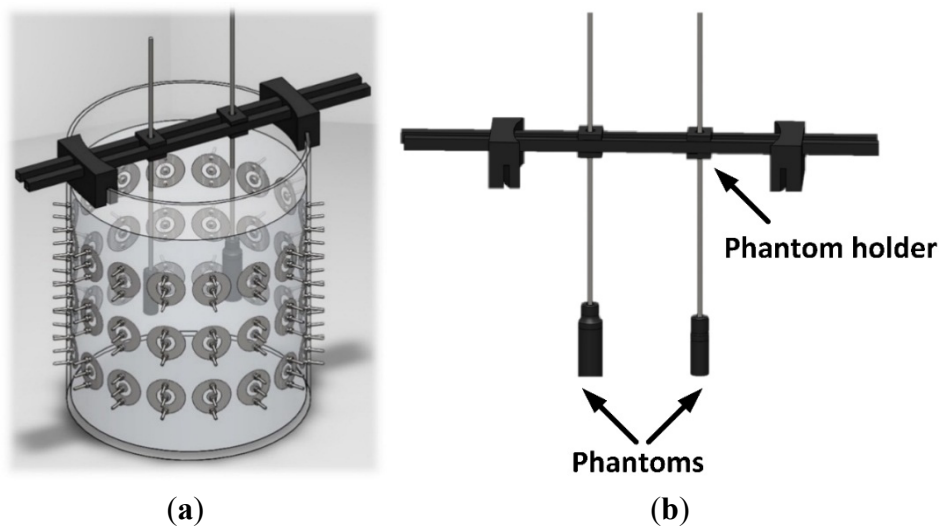

**Figure 12.** Constructed tank phantom (**a**) with phantom holder and phantoms (**b**).

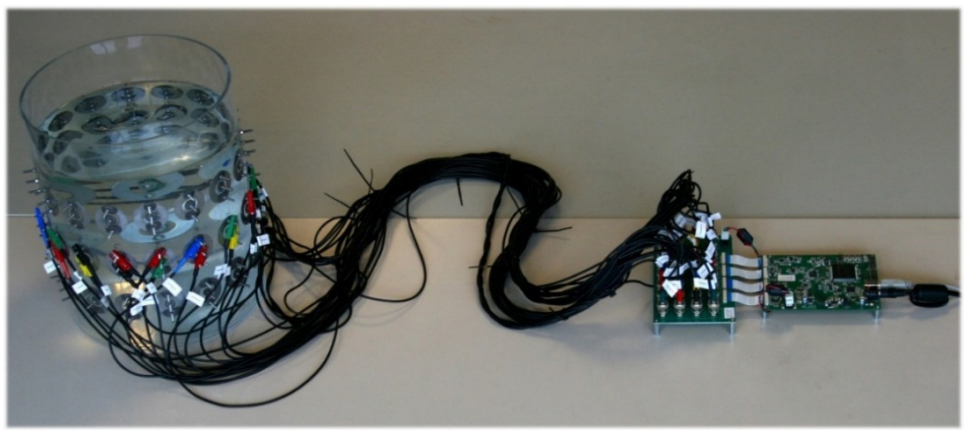

**Figure 13.** Measurement setup for detecting impedance changes inside a tank.

To analyze the signal quality, impedance measurements with 16 electrodes around the homogeneously salt water-filled tank are carried out. During this measurement the phantoms are not present inside the tank.

Subsequently the reciprocity accuracies ($RA_i$) of all measurement channels ($i = 1\ldots208$) are calculated via Equation (1), where ($\bar{z}_i$) is the mean value of 75 measurements and $\bar{z}_{r(i)}$ is the mean impedance value of the reciprocal measurement channel [20]:

$$RA_i = \left(1 - \frac{|\bar{z}_i - \bar{z}_{r(i)}|}{|\bar{z}_i|}\right) \cdot 100\ \%. \tag{1}$$

Figure 14 shows the measured reciprocity accuracies (**a**) and magnitudes of the transfer impedances between adjacent electrodes (**b**). The used sinusoidal excitation current has an amplitude of 5 mA with a frequency of 48.825 kHz. The low relative reciprocity accuracy of some channels in the first plot occurs because of the small impedance magnitudes using these channels, as can be seen in Figure 14b.

In these multiplexer configurations, the influence of the absolute reciprocity is significant, although the absolute reciprocity deviation of the system is less than ±70 mΩ. A common workaround for the problem of impedance magnitudes being too small is described in [21] and will be used subsequently.

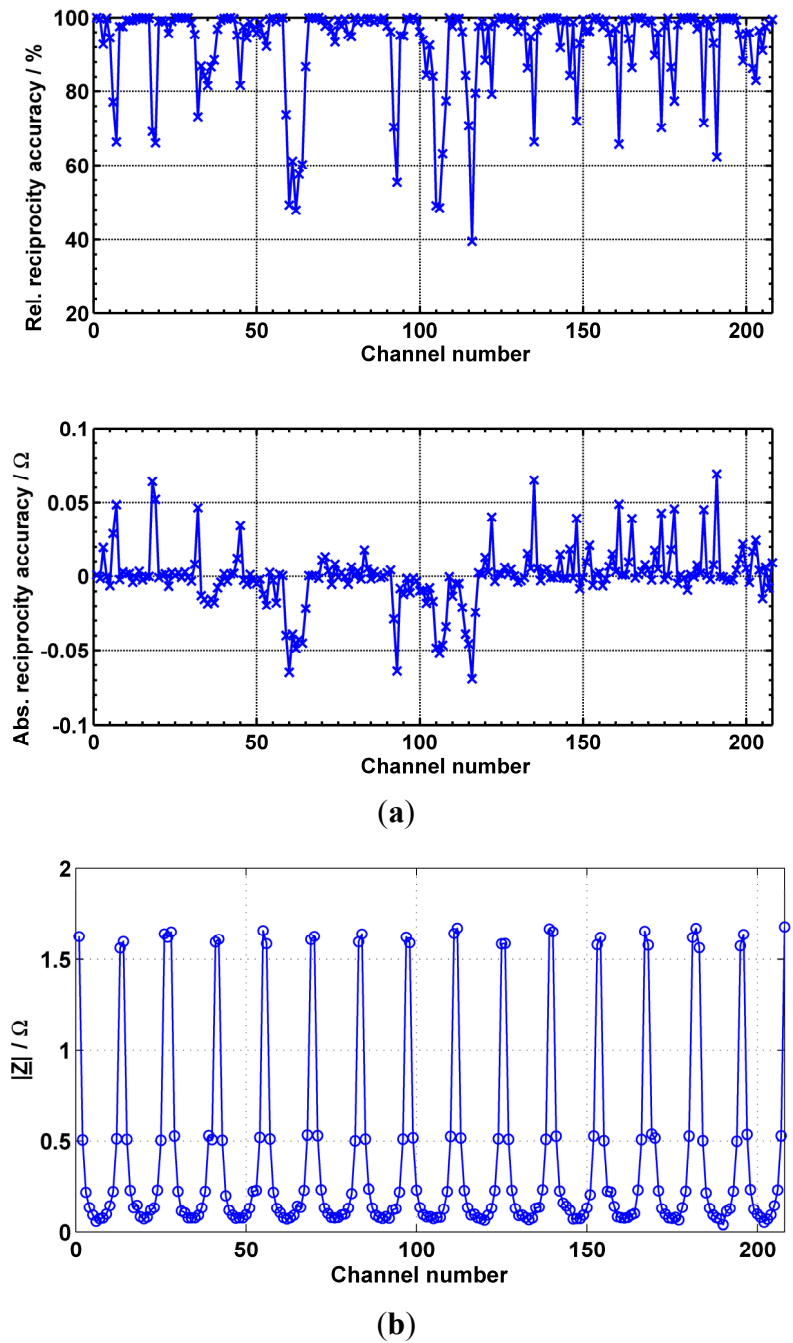

**Figure 14.** Reciprocity accuracies, measured with 16 electrodes around the tank. The sinusoidal excitation current had a frequency of 48.825 kHz and an amplitude of 5 mA. (**a**) Relative and absolute reciprocity accuracy of the measurement channels; (**b**) magnitudes of the measured transfer impedances over the different measurement channels.

For the reconstruction of the following measurements EIDORS in combination with NETGEN [22] and the GREIT algorithm [23] is used.

In Figure 15a the principal setup for the differential image acquisition is shown. Both phantoms are cylinders with diameters of 15 mm and 20 mm and heights are 30 mm and 50 mm, respectively, consisting of non-conductive ABS (Acrylonitrile Butadiene Styrene). Figure 15b shows the reconstructed difference image, where both phantoms can be identified. Also, the different sizes of the phantoms are distinguishable. In the reconstructed image, higher conductances are represented by bright areas and lower conductance by dark fields.

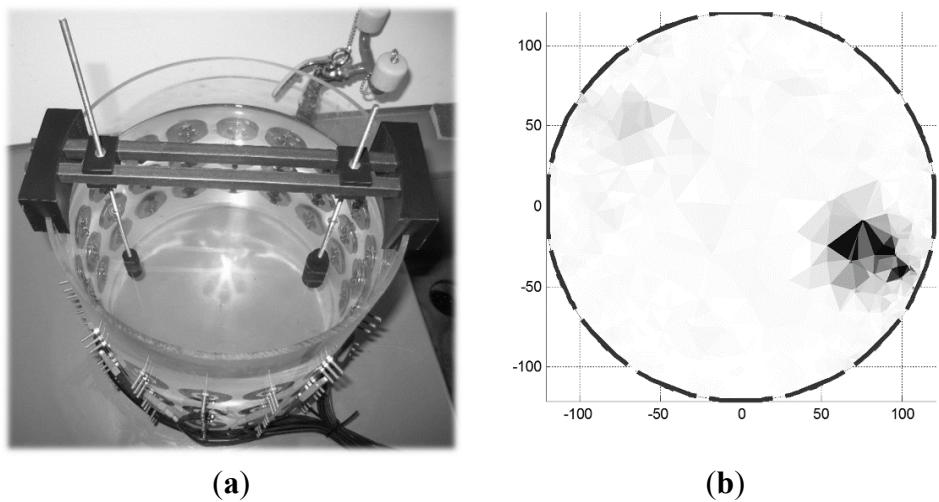

(**a**) (**b**)

**Figure 15.** Measurement and image reconstruction of ABS plastic phantoms in a homogeneous fluid, using a tank phantom with 16 connected electrodes. (**a**) Principle measurement setup for image acquisition with the tank phantom; (**b**) reconstruction result with both detected phantoms.

*3.3. Thorax Measurements*

After using tanks and phantoms to prove the characteristic of the EIT system, the next measurements are taken from a human thorax of a healthy male subject using standard silver chloride electrodes (Kendall Medi-Trace Mini 100 Snap Electrode). The aim of this measurement is the surveillance of the patient's respiration cycles. For the reconstruction a finite element method (FEM)-based forward model is used, which was derived from a computed tomography (CT) image [24]. It consists of 100,000 elements and is generated with EIDORS. For these first measurements only, the magnitudes of the impedances were used to reconstruct images. The excitation current had an amplitude of 5 mA and a frequency of 48.825 kHz. Data were acquired with an image rate of 4 IPS for a duration of 30 s over 208 channels. To enable good results, the healthy subject was requested to inhale and exhale very slowly. To improve the measurement further, the protocol was changed. Instead of measuring voltages between neighbored electrodes, a distance of seven electrodes was chosen according to the outcomes of [21].

The reconstruction results of the conductivity distribution is visualized in Figure 16 with time steps of 250 ms. The black parts of the images represent areas whose conductivity corresponds to the reference frame. Magenta areas have a decreased resistance, whereas white areas have the highest resistance.

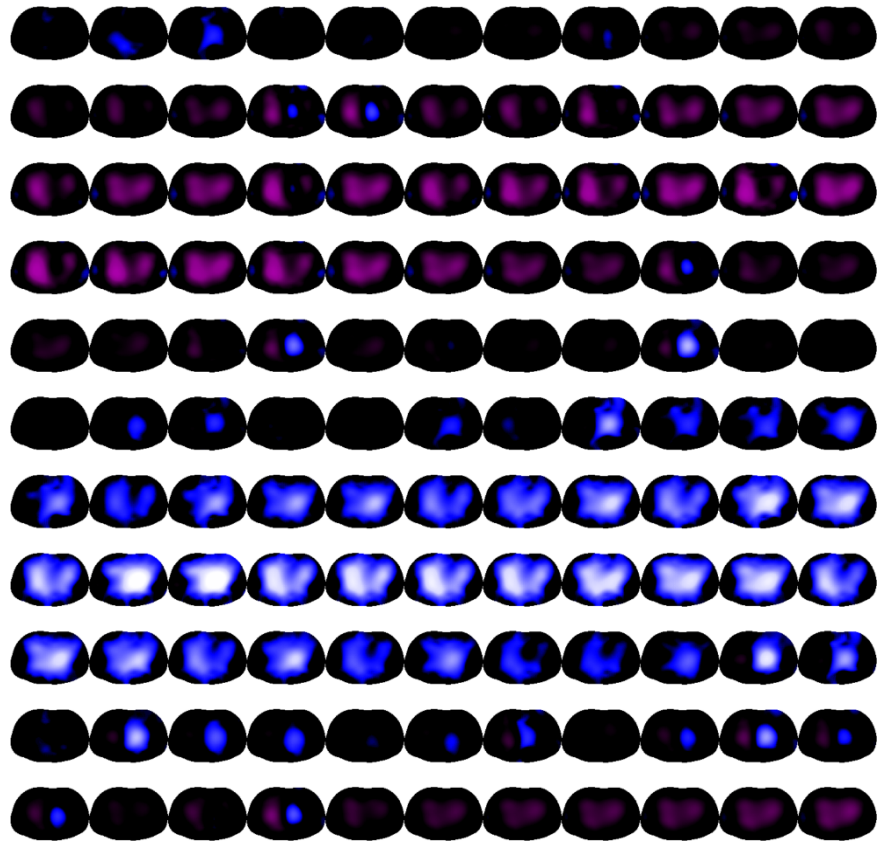

**Figure 16.** Visualization of reconstructed measurement results of a human thorax during one respiration cycle. The subject started exhaling (magenta). The duration of the measurement was 30 s with a frame rate of 4 FPS.

In Figure 17 the mean impedances averaged over all channels are plotted over the frame number and the time, respectively. The full breathing cycle, starting with expiration, which corresponds to lower impedance, is visible.

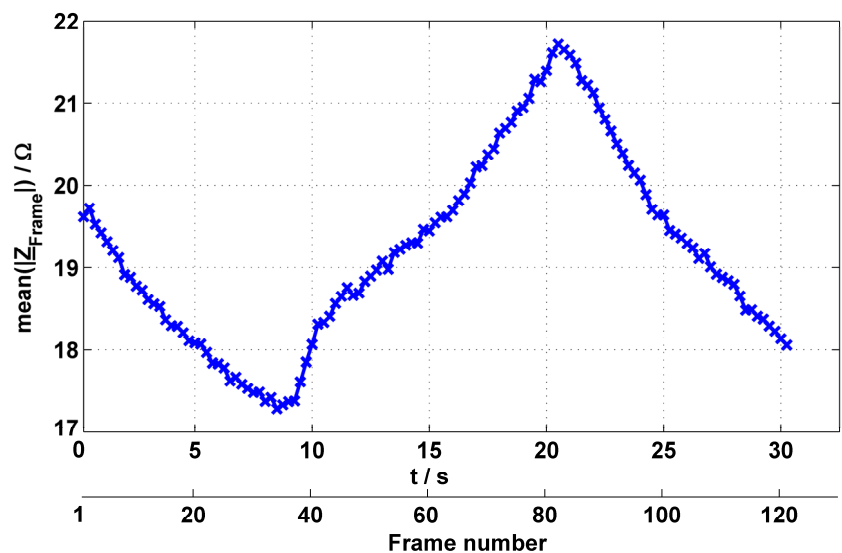

**Figure 17.** Mean impedances of each measurement frame during one breathing cycle.

## 4. Summary and Outlook

This work describes an FPGA-based multi-frequency EIT system for time-resolved measurements of conductivity distributions in living tissue. It is shown that the system acquires complex impedance values very precisely. To demonstrate the characteristics of the system, it was verified against an impedance phantom and real physiological measurements were carried out.

Besides adjustable excitation currents of up to 5 mA, the system can make use of different excitation signal forms such as sinusoidal, chirp, or rectangular with a frequency range of up to 500 kHz. Impedance measurements can be carried out with up to 16 multiplexed current and voltage channels each.

It was demonstrated that the achieved frame rate is sufficient to resolve the slow breathing of a subject. By using the phase information of the impedances in the reconstruction algorithm, the results can be improved in the future. To analyze faster events such as heart rate-related quantities, the frame rate has to be increased. This could be realized by executing isochronal voltage measurements. Since the first experiments with the hardware showed that the handling of up to 32 measurement cables is uncomfortable, the system–patient interface should be improved.

## Acknowledgments

This publication is a result of the ongoing research within the LUMEN research group, which is funded by the German Federal Ministry of Education and Research (BMBF, FKZ 13EZ1140A/B). The authors would also like to thank Analog Devices, Würth Elektronik, and Linear Technology for their support in terms of free samples during the development process.

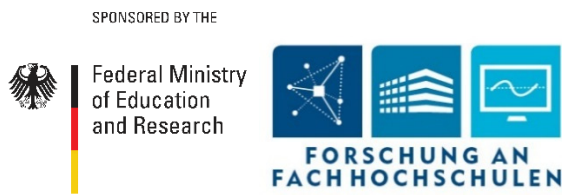

## Author Contributions

Roman Kusche contributed to the hardware and software development and took care of most of the writing. Gunther Ardelt and Ankit Malhotra contributed to the tests and measurements, the simulations and the interpretation of the results. Martin Ryschka guided the entire process and reviewed all measurement results and made corrections throughout the manuscript. Paula Klimach reviewed the simulation and measurement results and made corrections throughout the manuscript. Steffen Kaufmann developed the system architecture, hardware and software. All authors contributed to the manuscript preparation and approved the final manuscript.

## Conflicts of Interest

The authors declare no conflict of interest.